\documentclass[aps,prl,reprint,groupedaddress,nofootinbib]{revtex4-2}

\usepackage{epsfig}
\usepackage{graphicx}
\usepackage[figuresright]{rotating}

\usepackage{amsfonts}
\usepackage{amsmath}
\usepackage{amssymb}
\usepackage{wasysym}
\PassOptionsToPackage{normalem}{ulem}
\usepackage{ulem}
\usepackage[unicode=true,pdfusetitle,
bookmarks=true,bookmarksnumbered=false,bookmarksopen=false,
breaklinks=false,pdfborder={0 0 1},backref=false,colorlinks=false]
{hyperref}
\hypersetup{colorlinks,linkcolor=blue,citecolor=blue,urlcolor=blue}

\makeatletter

\usepackage{xcolor}
\usepackage{pdfcolmk}
\usepackage{wasysym}

\usepackage{txfonts}

\makeatother

\hypersetup{
	pdfborder={0 0 0},
	pdfnewwindow=true,      % links in new window
	colorlinks=true,       % false: boxed links; true: colored links
	linkcolor=blue,          % color of internal links (change box color with linkbordercolor)
	citecolor=blue,        % color of links to bibliography
	filecolor=blue,      % color of file links
	urlcolor=blue,          % color of external links
}
\usepackage{graphicx} % Include figure files
\usepackage{amsmath} % use \text
\usepackage[version=4]{mhchem} % chemical formula

\newcommand{\Eq}{Eq.}

\newcommand{\Fig}{Fig.}

\renewcommand{\vec}[1]{\boldsymbol{#1}}

\newcommand{\mi}{\mathrm{i}}

\usepackage{ulem} % strikeout outside math mode
\usepackage{cancel} % strikeout in math mode

\newcommand{\del}[1]{\ifmmode \textcolor{red}{\xcancel{#1}} 
\else \textcolor{red}{\sout{#1}} \fi} % delete text
\newcommand{\rev}[2]{\ifmmode \textcolor{red}{\xcancel{#1}} \textcolor{red}{#2} 
	\else \textcolor{red}{\sout{#1}} \textcolor{red}{#2} \fi} % revise text

\newif\ifdraft
\draftfalse

\newif{\ifshowcomments}
\showcommentstrue
\begin{document}

\title{Universal excitonic superexchange in spin-orbit-coupled Mott insulators}

\author{Chao-Kai Li}
\affiliation{Department of Physics and HKU-UCAS Joint Institute 
	for Theoretical and Computational Physics at Hong Kong, 
	The University of Hong Kong, Hong Kong, China}
\affiliation{The University of Hong Kong Shenzhen Institute of Research and Innovation, Shenzhen 518057, China}
\author{Gang Chen}
\email{gangchen@hku.hk}
\affiliation{Department of Physics and HKU-UCAS Joint Institute 
	for Theoretical and Computational Physics at Hong Kong, 
	The University of Hong Kong, Hong Kong, China}
\affiliation{The University of Hong Kong Shenzhen Institute of Research and Innovation, Shenzhen 518057, China}

\date{\today}

\begin{abstract}
We point out the universal presence of the excitonic superexchange in 
spin-orbit-coupled Mott insulators. 
It is observed that, the restriction to the lowest spin-orbit-entangled ``$J$'' states
may sometimes be insufficient to characterize the  
microscopic physics, and the virtual excitonic processes via the upper ``$J$'' states
provide an important correction to the superexchange. We illustrate  
this excitonic superexchange from a two-dimensional $5d$ iridate Sr$_2$IrO$_4$
and explain its physical consequences such as the orbital-like coupling to the
external magnetic flux and the nonlinear magnetic susceptibility.
The universal presence of the excitonic superexchange in other spin-orbit-coupled 
Mott insulators such as $3d$ Co-based Kitaev magnets and even $f$ electron 
rare-earth magnets is further discussed. 
\end{abstract}

% discuss f electron , crystal feild levels
% discuss 

% simplify equations, shorten equations, ...

\maketitle

\emph{Introduction.}---There has been a great interest in the field of 
spin-orbit-coupled correlated materials~\cite{Witczak_Krempa_2014}, ranging 
from the correlated spin-orbit-coupled metals or semimetals
to the spin-orbit-coupled Mott insulators. 
The latter covers the $4d$/$5d$ magnets like Kitaev materials, 
iridates, osmates~\cite{PhysRevB.78.094403,PhysRevLett.102.017205,PhysRevB.82.174440,PhysRevLett.105.027204,PhysRevB.84.094420,PhysRevB.91.219903,PhysRevLett.110.097204,PhysRevLett.112.077204,PhysRevB.91.241110,Rau_2016,PhysRevB.90.041112,trebst2017kitaev,Hermanns_2018}, 
$4f$ rare-earth magnets~\cite{Rau_2019,PhysRevB.94.035107,PhysRevResearch.2.043013,PhysRevB.95.041106,PhysRevB.94.201114,PhysRevB.94.075146,PhysRevLett.116.257204}, 
or even $3d$ Co-based Kitaev magnets
and Ni-based diamond antiferromagnets that are of some recent interest~\cite{Liu_2020,das2021realize,S0217979221300061,PhysRevB.97.014407,PhysRevB.97.014408,PhysRevMaterials.2.034404,PhysRevB.96.020412,PhysRevB.100.045103,PhysRevLett.120.057201,PhysRevB.100.140408}.  
In these materials, the local moments are formed by the 
spin-orbit-entangled ``$J$'' or ``$J_{\text{eff}}$'' states.
In the study of the superexchange interactions between the 
local moments, the prevailing assumption was to consider the pairwise
superexchange interaction between the neighboring local moments
from the lowest $J$ states, and the treatment was to carry out
the perturbation theory of the extended Hubbard model. 
The result from this assumption and treatment were 
somewhat successful, especially in the 
systems where the lowest $J$ states are very well separated from the 
excited or upper $J$ states and at the same time the Mott gap is large. 

In many of these spin-orbit-coupled Mott insulators, the spin-orbit coupling is not 
really a dominant energy scale especially in $3d$ or $4d$ transition metal compounds, 
and thus the energy separation between the lowest $J$ states and the upper $J$ 
states is not quite large compared to the hoppings and magnetic interactions, 
though it may involve both the crystal field effect
and the spin-orbit coupling. In rare-earth magnets, this energy separation is defined 
by the crystal field and can sometimes be small compared to the exchange 
energy scale. Moreover, the famous Kitaev materials like
iridates and $\alpha$-RuCl$_3$ are not good insulator as the charge gaps 
in these materials are not quite large~\cite{PhysRevResearch.1.013014,PhysRevB.94.161106,PhysRevB.90.041112,PhysRevB.93.075144}. In this work, 
we explore the excitonic superexchange process that
involves the upper $J$ states virtually and provide an important and universal correction 
to the superexchange interaction in the spin-orbit-coupled Mott insulators. 
The excitonic process refers to the tunneling of the electron from the lowest 
$J$ states to the upper $J$ states between the neighboring sites.

What was our previous knowledge about the superexchange 
interaction of the spin-orbit-coupled Mott insulators? 
For the ${J=1/2}$ moments, due to the absence of the continuous
rotational symmetry, all possible symmetry-allowed pairwise exchange interactions,
including Heisenberg, Dzyaloshinskii-Moriya and pseudo-dipole interactions~\cite{book_Physics_of_Transition_Metal_Oxides_Springer},
appear in the exchange matrix and are likely to be equally important. Moreover,
the diagonal entry of the exchange matrix from the Heisenberg and the 
pseudo-dipole interactions can yield the Kitaev~\cite{PhysRevLett.102.017205} or Kitaev-like anisotropic compass interaction~\cite{PhysRevLett.102.017205,PhysRevB.78.094403}.
For the larger $J$ moments (e.g. ${J=3/2}$ or $2$), 
due to the larger physical Hilbert space
and the strong accessibility via the spin-orbit entanglement, 
the pairwise interaction contains the high-order multipole 
interactions beyond the dipole moment 
interactions~\cite{PhysRevB.82.174440,PhysRevB.84.094420,PhysRevB.91.219903}. 
For the special quenched $J$ moments with ${J=0}$, 
the coupling with the excited $J$ states could lead to an
exciton Bose-Einstein condensation with 
magnetism~\cite{PhysRevLett.111.197201,PhysRevB.100.045103}.  
In this paper, we show that, for the unquenched local moment $J$,
the excitonic process to the excited $J$ states appears {\sl virtually}  
and renders an important correction of the 
superexchange interaction between the local moments, especially
in the presence of external magnetic field.

% pseudo-dipole 
 
% outline the results . 

% We begin by briefly reviewing the eigenstates of an atom with $ d^{5} $ configuration under the influence of the octahedral crystal field. The five degenerate $ d $ orbitals of an isolated atom will be split into two groups of orbitals with different energies. One group, called the $ t_{2g} $ orbitals, contains the orbitals $ d_{yz} $, $ d_{zx} $, and $ d_{xy} $. The other group, call the $ e_{g} $ orbitals, contains the orbitals $ d_{x^2-y^2} $ and $ d_{z^2} $.

\begin{figure}[b]
	\includegraphics[width=8cm]{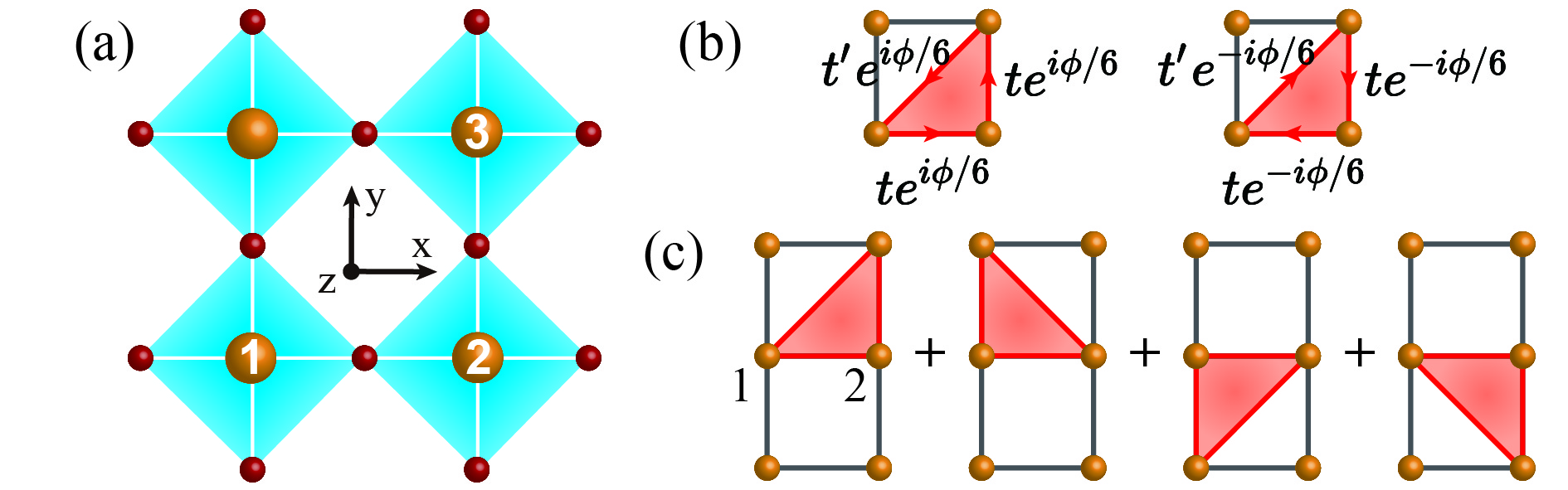}
	\caption{(a) The square lattice of $ d^{5} $ transition metals (orange balls) with corner sharing octahedra of oxygen ligands (red balls). Only the in-plane oxygen atoms are shown. (b) The Peierls phases attached to the hopping paths, where $ t $ ($ t^{\prime} $) is some nearest-neighbor (next-nearest-neighbor) hopping parameter. (c) For the spin interaction between site 1 and site 2, there are four types of hopping path for the virtual excitonic process, represented by the red triangles.}
	\label{fig: model}
\end{figure}

\emph{Atomic eigenstates.}---Because of the universality and the broad 
applicability of this virtual excitonic process, we deliver our theory via the 
simplest ${J=1/2}$ local moment for the $4d^5$ or $5d^5$ electron 
configuration of the Ir$^{4+}$ or Ru$^{3+}$ ion in the octahedral crystal 
field environment. This occurs for example in many iridates such as Sr$_2$IrO$_4$. 
The five degenerate atomic $ d $ orbitals split into the $ t_{2g} $ and the $ e_{g} $ manifolds, with the former having lower energy. In the case of a large crystal field, the energy gap is large, leading to a low-spin $ d^{5} $ configuration. There is one hole in the $ t_{2g} $ manifold which consists of orbitals $ d_{yz} $, $ d_{zx} $, and $ d_{xy} $. The $ t_{2g} $ manifold has an effective orbital angular momentum 
${ l_{\text{eff}} = 1 }$~\cite{Abragam1970}. Because there is a single hole in the atomic ground state, we will use the hole representation in this paper. The atomic Hamiltonian in the $ t_{2g} $ subspace at site $ i $ is
\begin{equation}\label{eq: atomic Hamiltonian}
H_{0i}=\frac{U}{2}\sum_{\substack{mm^{\prime}\sigma\sigma^{\prime}}
}\!\!\!\!\raisebox{4pt}{$^\prime$}n_{im\sigma}^{(e)}n_{im^{\prime}\sigma^{\prime}}^{(e)}-\lambda\sum_{mm^{\prime}\sigma\sigma^{\prime}}\vec l_{mm^{\prime}}\cdot\vec s_{\sigma\sigma^{\prime}}c_{im\sigma}^{\dagger}c_{im^{\prime}\sigma^{\prime}}^{},
\end{equation}
where $ c_{im\sigma}^{\dagger} $ ($ c_{im\sigma} $) is the hole creation (annihilation) operator with orbital ${ m = 1, 2, 3 }$ and spin ${ \sigma = \uparrow, \downarrow }$ at site $ i $. The first term is the Hubbard interaction with strength ${ U > 0 }$. The prime of the summation means that the term with ${ (m\sigma)=(m^{\prime}\sigma^{\prime}) }$ is excluded. Note that a hole creation (annihilation) operator is an electron annihilation (creation) operator. Thus, the electron occupation number ${ n_{im\sigma}^{(e)}=1-n_{im\sigma} }$ in the hole representation. The Hund's interactions are relatively small and hence neglected. The last term is the spin-orbit interaction with strength ${\lambda > 0}$. In the hole representation, the sign before $ \lambda $ is negative. The orbital index ${ m = 1, 2, 3 }$ corresponds to the three $ t_{2g} $ orbitals $ d_{yz} $, $ d_{zx} $, and $ d_{xy} $, respectively. In this index ordering, the orbital angular momentum matrix elements $ (l_{m})_{m^{\prime}m^{\prime\prime}}=\mi\epsilon_{mm^{\prime}m^{\prime\prime}} $, in which $ l_{1,2,3} $ refers to $ l_{x,y,z} $, respectively. The sign is opposite to the matrix elements of the genuine ${ l = 1 }$ orbitals~\cite{Abragam1970}. The spin angular momentum ${ \vec s=\vec{\sigma}/2 }$, where ${ \vec{\sigma}=(\sigma_{x},\sigma_{y},\sigma_{z}) }$ is the vector of Pauli matrices.

The eigenstates of the SOC term are two ${ j_{\text{eff}} = 1/2 }$ states with energy $ {-\lambda} $, and four ${ j_{\text{eff}} = 3/2 }$ states with energy $ \lambda / 2 $. In the atomic limit, the hole lies on the two-fold degenerate ${ j_{\text{eff}} = 1/2 }$ level, making the ion an effective spin-1/2 system. As long as the hoppings between different sites remain relatively weak as compared with the on-site Coulomb repulsion, the system remains a spin-orbit-coupled Mott insulator.

\emph{Exchange interaction.}---The interactions between the effective spins has their origin in the virtual hopping processes. When the hopping terms between different sites are taken into account, there are virtual intermediate states that contain more than one hole on the same ion site, at the cost of Coulomb repulsion energy. The intermediate states are subject to the Pauli exclusion principle, which prevents two holes with the same spin from occupying the same state. Therefore, the resulting energy correction will be dependent on the spin configurations of the initial and finial states. The overall effect is the development of an effective interaction between neighboring spins, which is called exchange interaction.

In general, the hopping processes can be described by the Hamiltonian
\begin{equation}\label{eq: TB model}
	T=\sum_{ij\sigma}C_{i\sigma}^{\dagger}h_{ij}C_{j\sigma},
\end{equation}
in which ${ h_{ij} = h_{ji}^{\dagger} }$ is the matrix for the hopping from site $ j $ to site $ i $, and $ C_{i\sigma}^{\dagger}=[c_{i,yz,\sigma}^{\dagger},c_{i,zx,\sigma}^{\dagger},c_{i,xy,\sigma}^{\dagger}] $ is a row vector of the $ t_{2g} $ hole creation operators at site $ i $ with spin $ \sigma $. The Hamiltonian of the entire system is ${ H=H_{0}+T }$, where ${ H_{0}=\sum_{i}H_{0i} }$ is the summation of the atomic Hamiltonian \eqref{eq: atomic Hamiltonian} on each site. For a Mott insulator, $ H_0 $ dominates $ T $, so the latter can be treated by perturbation theory. The matrix element of the effective Hamiltonian up to third order is
\begin{equation}\label{eq: perturbation expansion}
(H_{\text{eff}})_{mn}=(H_{0})_{mn}+\sum_{\alpha}\frac{T_{m\alpha}T_{\alpha n}}{\omega_{0\alpha}}+\sum_{\alpha\beta}\frac{T_{m\alpha}T_{\alpha\beta}T_{\beta n}}{\omega_{0\alpha}\omega_{0\beta}}
\end{equation}
where the Latin indices $ m $ and $ n $ refer to the ground states of the unperturbed Hamiltonian 
$ H_0 $, and the Greek indices $ \alpha $ and $ \beta $ refer to the excited states. $ \omega_{0\alpha} $ is the difference between the unperturbed energies of the ground state 
$ 0 $ and the excited state $ \alpha $.

We begin by a simple model shown in \Fig~\ref{fig: model}(a), in which the $ d^5 $ octahedrally coordinated transition metals arrange in a two-dimensional square lattice, the same as the \ce{CuO2} plane in cuprates.

\begin{figure}[t]
	\includegraphics[width=8cm]{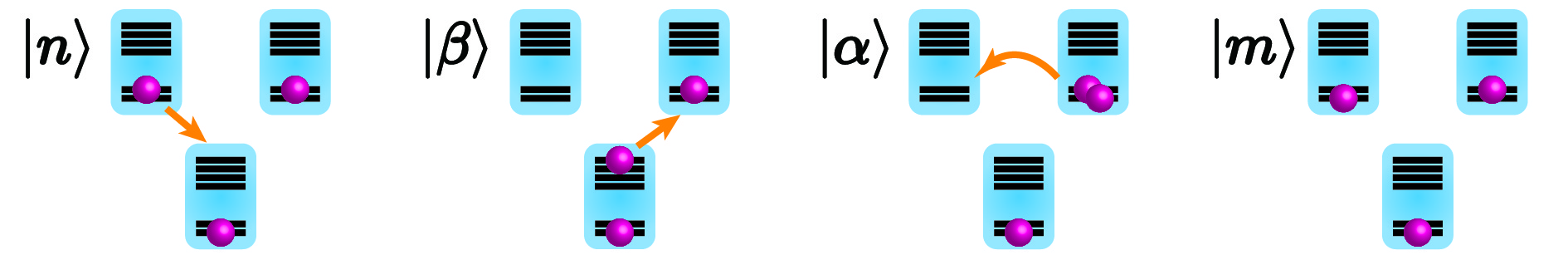}
	\caption{An example of virtual excitonic process. 
	Each blue rectangle represents a $d^{5}$ transition metal ion, 
	which has a two-fold degenerate ${ j_{\text{eff}} = 1/2}$ level 
	and a four-fold degenerate ${j_{\text{eff}} = 3/2}$ level. 
	${ |n\rangle }$ and ${ |m\rangle }$ are the initial state and the final state, 
	respectively. ${|\beta\rangle }$ and ${|\alpha\rangle }$ 
	are two intermediate excited states.}
	\label{fig: exciton_process_example}
\end{figure}

By symmetry considerations, the hopping matrices for $ 2 \rightarrow 1 $ and $ 3 \rightarrow 1 $ in \Fig~\ref{fig: model}(a) have the form
\begin{equation}
h_{12}=\left[\begin{array}{ccc}
	t_{\delta} & 0 & 0\\
	0 & t_{\pi} & 0\\
	0 & 0 & t_{\pi}
\end{array}\right]
\quad\text{and}\quad
h_{13}=\left[\begin{array}{ccc}
	\frac{t_{\pi}^{\prime}+t_{\delta}^{\prime}}{2} & \frac{t_{\pi}^{\prime}-t_{\delta}^{\prime}}{2} & 0\\
	\frac{t_{\pi}^{\prime}-t_{\delta}^{\prime}}{2} & \frac{t_{\pi}^{\prime}+t_{\delta}^{\prime}}{2} & 0\\
	0 & 0 & \frac{3t_{\sigma}^{\prime}+t_{\delta}^{\prime}}{4}
\end{array}\right],
\end{equation}
respectively. Their subscripts hint that the hoppings are similar to the corresponding $ dd $-type Slater-Koster parameters, although usually the ligand-mediated indirect hopping is the dominant mechanism. All the other hopping matrices up to second nearest neighbor can be derived by symmetry. The hopping terms for further neighbors are neglected.

Traditionally, the excited states are limited to the ones that have two holes lying on the same $ j_{\text{eff}} = 1/2 $ manifold, similar to the one-band Hubbard model~\cite{PhysRevLett.106.136402}. With this restriction, only the second-order perturbation term in \Eq~\eqref{eq: perturbation expansion} contributes to the nearest-neighbor spin-spin interaction. By the definition of the spin operator at site $ i $ as ${ (\vec S_{i})_{ss^{\prime}}=c_{is}^{\dagger}\vec{\sigma}_{ss^{\prime}}c_{is^{\prime}}^{}/2 }$ where $ s $ and $ s^{\prime} $ refer to the two $ j_{\text{eff}} = 1/2 $ states, the effective Hamiltonian for the spin interaction between site 1 and site 2 is derived as
\begin{equation}\label{eq: traditional exchange}
H_{12}=\frac{4(2t_{\pi}+t_{\delta})^{2}}{9U}\vec S_{1}\cdot\vec S_{2}.
\end{equation}
This is similar to the result for the one-band Hubbard model. If ${ t_{\pi} = t_{\delta} = t }$, 
then the coefficient $ 4t^2 / U $ is the expected antiferromagnetic 
Heisenberg exchange interaction parameter.

\emph{Virtual excitonic process.}---In spin-orbit-coupled Mott insulator the ${ j_{\text{eff}} = 1/2 }$ level is not well separated from the ${ j_{\text{eff}} = 3/2 }$ level, as can be seen in the typical spin-orbit-coupled Mott insulators \ce{Sr2IrO4}~\cite{Kim2008}, \ce{Na2IrO3}~\cite{Comin2012}, and \ce{$\alpha$-RuCl3}~\cite{PhysRevB.90.041112}. The existence of the nearby ${ j_{\text{eff}} = 3/2 }$ level will have an impact on the spin exchange interaction.

After taking the $ {j_{\text{eff}} = 3/2 }$ level into consideration, the intermediate excited states represented by the indices $ \alpha $ and $ \beta $ in \Eq~\eqref{eq: perturbation expansion} not only contain those in which two holes occupy the ${ j_{\text{eff}} = 1/2} $ level, but also those in which one hole occupies the ${ j_{\text{eff}} = 1/2 }$ level and the other hole occupies the $ {j_{\text{eff}} = 3/2 }$ level \footnote{The intermediate states in which two holes occupy the ${ j_{\text{eff}} = 3/2 }$ level appear only in higher order terms in the perturbative expansion.}. The process involving the latter type of excited states is called the virtual excitonic process, because it is as if one hole is excited from the $ {j_{\text{eff}} = 1/2 }$ level to the ${ j_{\text{eff}} = 3/2} $ level, leaving an electron behind, and the virtual pair of the electron and the excited hole forms a virtual exciton.

Allowing the virtual excitonic process in the perturbation expansion \eqref{eq: perturbation expansion}, it is found that the correction to the two-spin exchange interaction comes from the third order term, which involves the participation of a third ion, as schematically shown in \Fig~\ref{fig: exciton_process_example}.

The involvement of a third ion has significant consequences when an out-of-plane magnetic field is applied. There will be additional Peierls phases in the hopping matrices. The accumulated Peierls phase for a closed loop should equal to $ 2\pi $ times the magnetic flux through the loop area divided by the flux quantum $ hc/e $. Assuming that the accumulated Peierls phase for a plaquette is $ \phi $, it is convenient to use the gauge indicated in \Fig~\ref{fig: model}(b). Note that the second order interaction \eqref{eq: traditional exchange} is independent of the magnetic field because the areas of the hopping paths ${ 1 \rightarrow 2 \rightarrow 1 }$ and ${ 2 \rightarrow 1 \rightarrow 2 }$ 
are zero.

For a square lattice up to second-nearest-neighbor hopping, \Fig~\ref{fig: model}(c) shows the four hopping paths at the third perturbation order that contribute to the exchange interaction between site 1 and site 2. The resulting correction term is found to be
\begin{equation}\label{eq: exciton correction}
H_{12}^{\prime}=\frac{16(\lambda+U)(2t_{\pi}+t_{\delta})(t_{\delta}-t_{\pi})(3t_{\sigma}^{\prime}-2t_{\pi}^{\prime}-t_{\delta}^{\prime})}{9U(3\lambda+2U)^{2}}\cos\frac{\phi}{2}\vec S_{1}\cdot\vec S_{2}
\end{equation}
In sharp contrast to \Eq~\eqref{eq: traditional exchange}, the correction term is dependent on the magnetic field. This is because the area of the third order hopping processes like $ 1 \rightarrow 2 \rightarrow 3 \rightarrow 1 $ is nonzero, leading to a nonvanishing magnetic flux. Such field dependence is fundamentally different from the Zeeman interaction, because it is essentially coupled to the magnetic flux, which is an orbital effect. The above analysis can be easily generalized to different lattice models, so the effect should be ubiquitous in various spin-orbit-coupled Mott insulators.

At the third order perturbation, there are also the ring exchange interactions among each group of three neighboring spins. For the spins 1, 2, and 3 in \Fig~\ref{fig: model}(a), the ring exchange interaction is
\begin{equation}\label{eq: ring exchange, no exciton}
H_{123}=-\frac{2\left(2t_{\pi}+t_{\delta}\right)^{2}\left(3t_{\sigma}^{\prime}+4t_{\pi}^{\prime}+5t_{\delta}^{\prime}\right)}{9U^{2}}\sin\frac{\phi}{2}\thinspace\vec S_{1}\cdot\left(\vec S_{2}\times\vec S_{3}\right).
\end{equation}
Because the mixed product of three spins breaks the time-reversal symmetry, the 3-spin ring exchange exists only when a magnetic field is applied. Hence, it must be field-dependent. The result \eqref{eq: ring exchange, no exciton} does not involve the virtual excitonic process. The hole only hops between the ${ j_{\text{eff}} = 1/2 }$ states of the three ions. The corrections by the virtual excitonic process comes at the fourth perturbation order, which engages a fourth ion to provide the ${ j_{\text{eff}} = 3/2}$ states. We expect that the correction terms have a field dependence of $ \sin\phi $, because the loop area is that of a plaquette, and the product $ \left[ \sin\phi\ \vec S_{1}\cdot\left(\vec S_{2}\times\vec S_{3}\right) \right] $ is time-reversal invariant.

\begin{figure}
	\includegraphics[width=8.2cm]{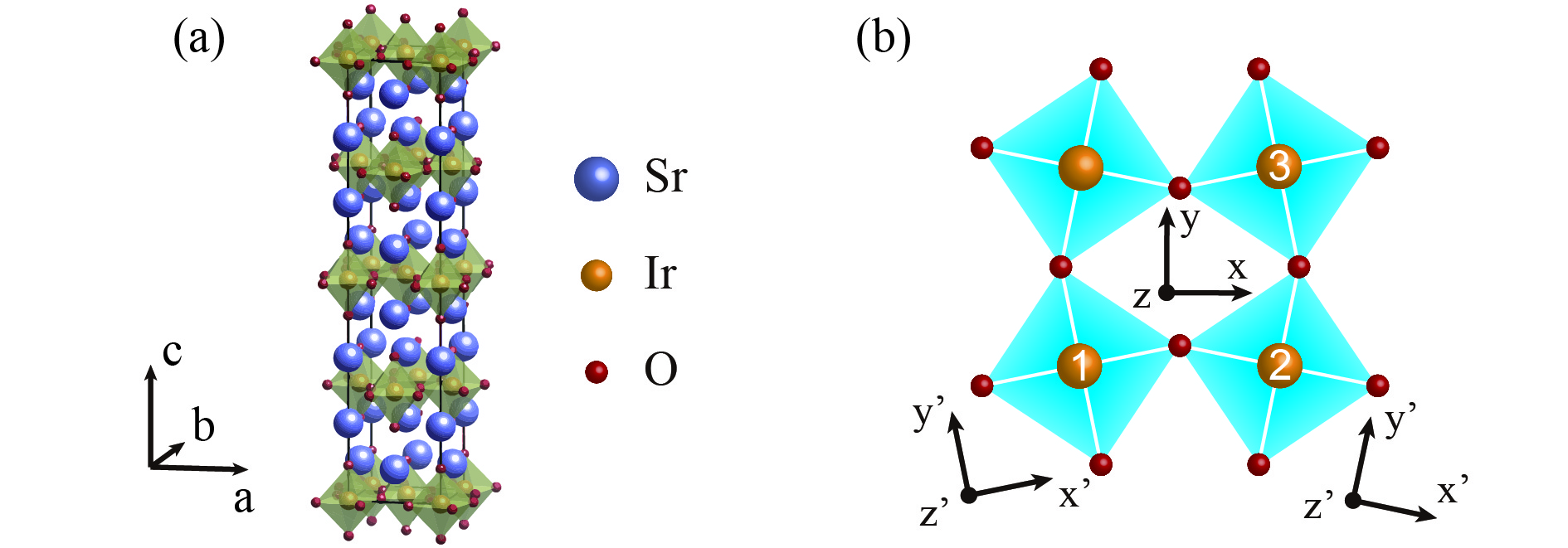}
	\caption{(a) The conventional unit cell of \ce{Sr2IrO4}. It is composed of four layers stacking along the $ c $ direction. (b) A \ce{IrO2} plane. The corner sharing \ce{IrO6} octahedra rotate alternately, resulting in two sets of local coordinate systems $ x^{\prime}y^{\prime}z^{\prime} $.}
	\label{fig: crystal_structure_Sr2IrO4}
\end{figure}

\emph{Application to \ce{Sr2IrO4}.}---Having demonstrated the basic concept of the virtual excitonic process and the results on a simple model, we take a look at its effects in real material. To demonstrate it, we use \ce{Sr2IrO4} as an example. 
Sr$_2$IrO$_4$ is a well-known spin-orbit-coupled insulator~\cite{PhysRevLett.102.017205,PhysRevLett.106.136402,PhysRevB.103.155147}. 
The electron configuration of the \ce{Ir^{4+}} ion is $ 5d^{5} $, the same as our assumption before. It consists of corner sharing \ce{IrO6} octahedra layers with intervening \ce{Sr} layers, see \Fig~\ref{fig: crystal_structure_Sr2IrO4}(a). The crystal structure of \ce{Sr2IrO4} is the \ce{Ca2MnO4} type with space group $ I4_{1}/acd $~\cite{Huang1994}. Compared to the above simple model, the complication is that the \ce{IrO6} octahedra rotate alternately from perfect alignment, as shown in \Fig~\ref{fig: crystal_structure_Sr2IrO4}(b).

The TB model without SOC is still given by the form of \Eq~\eqref{eq: TB model}. But we note that here the $ d $ orbitals in the subscripts are defined with respect to the local coordinate system aligned with the \ce{IrO6} octahedron at the corresponding site, as shown by the two sets of local coordinate systems $ x^{\prime}y^{\prime}z^{\prime} $ in \Fig~\ref{fig: crystal_structure_Sr2IrO4}(b). On the contrary, the spin directions are defined with respect to the global $ xyz $ axes in \Fig~\ref{fig: crystal_structure_Sr2IrO4}(b), because in the end we would like to have a spin interaction Hamiltonian expressed in the global coordinate system.

We perform the first-principles calculations to obtain the parameters in the Hubbard Hamiltonian. Wannier functions and the tight-binding model are constructed from the Kohn-Sham orbitals, and constrained random phase approximation is used to calculate the interaction parameter $ U $. See details in the Supplementary Material \href{http://stacks.iop.org/0295-5075/139/56001/mmedia}{\texttt{Supplementarymaterial.pdf}} (SM), which includes refs.~\cite{Marzari1997, Souza2001, Hohenberg1964, Kohn1965, QE-2009, QE-2017, Huang1994, Schlipf2015, Hamann2013, Aryasetiawan2011, Nakamura2021, Hinuma2017, Togo2018}.

Considering the crystal symmetry, the hole hopping matrices between the nearest-neighbor pair 12 and the second-nearest-neighbor pair 13 in \Fig~\ref{fig: crystal_structure_Sr2IrO4}(b) are
\begin{equation}
h_{12}=-\left[\begin{array}{ccc}
	-t_{3} & t_{4} & 0\\
	-t_{4} & -t_{1} & 0\\
	0 & 0 & -t_{2}
\end{array}\right]
\  \text{and}\  
h_{13}=-\left[\begin{array}{ccc}
	t_{3}^{\prime} & -t_{4}^{\prime} & 0\\
	-t_{4}^{\prime} & t_{2}^{\prime} & 0\\
	0 & 0 & -t_{1}^{\prime}
\end{array}\right]
\end{equation}
respectively. The overall minus signs are to emphasize that the hopping matrices for holes differ from those for electrons by a sign, while the latter is the direct result of the first-principles calculations. We note that the zero matrix elements in $ h_{12} $ and $ h_{13} $ are not strict because of the interlayer interactions. However, their magnitudes are around $ 0.1~\text{meV} $, much smaller than the other matrix elements. Thus, the smallness of these matrix elements reflects the weakness of the interlayer interactions. Other hopping matrices can be inferred by the crystal symmetry. The hopping parameters $ t_{1,2,3,4} $ and $ t_{1,2,3,4}^{\prime} $, the SOC strength $ \lambda $, and the Hubbard interaction parameter $ U $ are listed in Table~\ref{tab: parameters}.

By applying the perturbation theory, the spin interaction between the pair 12 in \Fig~\ref{fig: crystal_structure_Sr2IrO4}(b) is found to be
\begin{eqnarray}\label{eq: spin interation of Sr2IrO4}
H_{12}&=&J\vec S_{1}\cdot\vec S_{2}-D(\vec S_{1}\times\vec S_{2})_{z}+KS_{1}^{z}S_{2}^{z} \nonumber\\
&&-\left[J^{\prime}\vec S_{1}\cdot\vec S_{2}+D^{\prime}(\vec S_{1}\times\vec S_{2})_{z}-K^{\prime}S_{1}^{z}S_{2}^{z}\right]\bigg\}\cos\frac{\phi}{2}.
\end{eqnarray}
The analytical expressions for the coefficients are given in the SM, and their numerical values are listed in Table~\ref{tab: parameters}. The terms involving $ J^{\prime} $, $ D^{\prime} $, and $ K^{\prime} $ come from the virtual excitonic processes. The phase $ \phi=eBl^{2}/\hbar $, where $ B $ is the out-of-plane magnetic field, and $ l = 3.88~\mathring{\text{A}} $ is the distance between nearest-neighbor \ce{Ir} atoms. The order of magnitude $ \phi \sim 1 $ corresponds to a magnetic field $ B \sim 4.4 \times 10^{3}~\text{T} $.

\emph{Nonlinear susceptibility.}---The positive coefficient before $ S_{1}^{z}S_{2}^{z} $ leads to the easy-plane anisotropy. The in-plane magnetic structure is canted antiferromagnetic because of the competition between the Heisenberg and the Dzyaloshinskii-Moriya interaction (the first two terms of \Eq~\eqref{eq: spin interation of Sr2IrO4}). With the application of a magnetic field $ \vec{B} $ perpendicular to the \ce{IrO2} layers, the canted antiferromagnetic structure develops an out-of-plane component. We first neglect the virtual excitonic process corrections, then the classical energy per site is
\begin{equation}
E=2\left[J\vec S_{\text{A}}\cdot\vec S_{\text{B}}-D\left(\vec S_{\text{A}}\times\vec S_{\text{B}}\right)_{z}+KS_{\text{A}}^{z}S_{\text{B}}^{z}\right]-2\mu_{\text{B}}BS_{\text{A}}^{z},
\end{equation}
where $ \vec S_{\text{B}} $ and $ \vec S_{\text{B}} $ are the spins on the two sublattices, with the same $ z $ component, and $ \mu_{\text{B}} $ is the Bohr magneton. We have used the fact that the Land\'e $ g $-factor of the ${ j_{\text{eff}} = 1/2 }$ hole state is ${ -2 }$. By minimizing the energy, the tilting angle of the spins away from the $ xy $ plane can be obtained, and the perpendicular magnetic susceptibility is found to be
\begin{equation}
\label{eq: chi without correction}
\chi_{\perp}=\frac{N\mu_{\text{B}}^{2}}{\sqrt{J^{2}+D^{2}}+J+K},
\end{equation}
in which $ N $ is the site density. This result~\eqref{eq: chi without correction} has the same property as the mean-field result for the N\'eel antiferromagnetism that below the N\'eel temperature it depends neither on the temperature nor on the magnetic field.

Now we take into account the correction term from the virtual excitonic process. It amounts to the replacements ${ J\rightarrow J-J^{\prime}\cos(\phi/2)} $, ${ D\rightarrow D+D^{\prime}\cos(\phi/2) }$, and ${ K\rightarrow K+K^{\prime}\cos(\phi/2) }$. With Taylor expansion up to $ O(\phi^2) $, we find the magnetization
\begin{equation}
M=\chi_{\perp}^{(1)}B+\chi_{\perp}^{(3)}B^{3}+\cdots,
\end{equation}
where
\begin{equation}
\chi_{\perp}^{(1)}=\frac{N\mu_{\text{B}}^{2}}{\sqrt{(J-J^{\prime})^{2}+(D+D^{\prime})^{2}}+J-J^{\prime}+K+K^{\prime}}
\end{equation}
is linear susceptibility, and
\begin{equation}
\chi_{\perp}^{(3)}=-\frac{N\mu_{\text{B}}^{2}\left[\frac{(J-J^{\prime})J^{\prime}-(D+D^{\prime})D^{\prime}}{\sqrt{(J+J^{\prime})^{2}+(D-D^{\prime})^{2}}}+J^{\prime}-K^{\prime}\right]\left(\frac{el^{2}}{\hbar}\right)^{2}}{8\left[\sqrt{(J-J^{\prime})^{2}+(D+D^{\prime})^{2}}+J-J^{\prime}+K+K^{\prime}\right]^{2}}
\end{equation}
is the nonlinear susceptibility~\cite{Fujiki1981,Machida2010}. 
Its existence is the consequence of the virtual excitonic process.
This clarifies one important source of the nonlinear susceptibility
in these systems.

\begin{table}
	\begin{ruledtabular}
		\begin{tabular}{cccccccccc}
			$ t_{1} $ & $ t_{2} $ & $ t_{3} $ & $ t_{4} $ & $ t_{1}^{\prime} $ & $ t_{2}^{\prime}$ & $ t_{3}^{\prime} $ & $ t_{4}^{\prime} $ & $ \lambda $ & $ U $ \\
			$ 285 $ & $ 243 $ & $ 46.8 $ & $ 27.8 $ & $ 125 $ & $ 30.1 $ & $ 5.92 $ & $ 11.1 $ & $ 365 $ & 2232 \\
			\toprule
			$ J $ & $ D $ & $ K $ & $ J^{\prime} $ & $ D^{\prime} $ & $ K^{\prime}$ \\
			$ 65.2 $ & $ 12.7 $ & $ 1.23 $ & $ 3.51 $ & $ 0.895 $ & $ 0.237 $
		\end{tabular}
	\end{ruledtabular}
	\caption{The parameters of the Hubbard model of \ce{Sr2IrO4} (top), and the parameters of the spin interaction between site 1 and site 2 in \Fig~\ref{fig: crystal_structure_Sr2IrO4}(b) (bottom), in meV.}
	\label{tab: parameters}
\end{table}

% weak lambda soc, small U , weak crystal field gap, is the applicablity of excitonic superexchange 

\emph{Discussion.}---Here we discuss the applicability of the excitonic superexchange. 
We have shown that, the virtual excitonic process appears in the high order perturbation
theory of the multiple-band Hubbard model. A smaller spin-orbit-coupling and/or a weak
Hubbard interaction could enhance its contribution. Apparently, in many of these 
$4d/5d$ magnets, the spin-orbit coupling is not really the dominant energy scale, 
and the systems also behave like weak Mott insulators. The $4d$ magnet 
$\alpha$-\ce{RuCl3} is a good example of this kind. This material shows a
remarkable thermal Hall transport result in external magnetic fields, 
which is likely to be related to the gapped Kitaev spin liquid~\cite{Yokoi_2021,Kasahara_2018,Kasahara_20182}. 
More recently, it has been found that its thermal conductivity    
oscillates when an external magnetic field is applied~\cite{Czajka2021}. 
It is believed that the oscillation originates from the existence 
of the spinon Fermi surface of some kind. The orbital quantization 
of the spinons requires the coupling of the spinons with an internal
U(1) gauge field~\cite{PhysRevB.73.155115}. Such orbital effect 
is obviously not controlled via the Zeeman term. The virtual 
excitonic process and/or the strong charge fluctuation could 
provide the microscopic means to lock the internal U(1) gauge 
field with the external magnetic flux, and thereby generating 
the oscillatory behaviors. 

For the Co-based Kitaev materials~\cite{Liu_2020,das2021realize,S0217979221300061,PhysRevB.97.014407,PhysRevB.97.014408}, the Hubbard interaction
is large, but the spin-orbit coupling is weak. Thus, the excitonic
superexchange may still be important. 
For the rare-earth magnets, the crystal field enters as
an important energy scale. If the crystal field energy separation
between the ground states and the excited states is not large 
enough compared to the exchange energy scale, the excited 
states would be involved into the low-temperature magnetic 
physics. This upper-branch magnetism has been invoked for the pyrochlore magnet 
Tb$_2$Ti$_2$O$_7$~\cite{PhysRevB.84.140402,PhysRevLett.98.157204,Liu_2019,kadowaki2021spin}.

The concept could also be applied to systems such as metal organic frameworks, molecular magnets, and Moir\'e systems, which have large lattice constants. Another example is the spin-liquid candidate material $ 1T $-\ce{TaS2}~\cite{Law6996,PhysRevLett.121.046401,Butler2020,Wang2020,Li2022}, which enters into a charge-density-wave (CDW) phase at low temperature. In the CDW phase, each supercell has a localized unpaired electron which provides the spin. The neighboring spins are separated at a distance $ l $ larger than ten angstroms. Moreover, the orbitals $ d_{xy} $, $ d_{x^2-y^2} $, and $ d_{z^2} $ have similar energies~\cite{Rossnagel2006}, which facilitates the excitonic process. The required magnetic field $ B $ for the phase $ \phi \sim eBl^{2}/\hbar $ to be at the order of $ O(1) $ drops to hundreds of Tesla, making the effect of excitonic process more prominent.

\begin{acknowledgments}
\emph{Acknowledgments.}---The first-principles calculations for this work 
were performed on TianHe-2. We are thankful for the support from the 
National Supercomputing Center in Guangzhou (NSCC-GZ).
This work is supported by the National Science Foundation of China 
with Grant No. 92065203, the Ministry of Science and Technology 
of China with Grants No. 2018YFE0103200, by the Shanghai Municipal Science 
and Technology Major Project with Grant No. 2019SHZDZX01, 
and by the Research Grants Council of Hong Kong 
with General Research Fund Grant No. 17306520. 
\end{acknowledgments}

\bibliography{refs.bib}

\end{document}

\title{Supplementary Material for
	``Universal Excitionic Superexchange in Spin-orbit-coupled Mott Insulators''}

\author{Chao-Kai Li}
\affiliation{Department of Physics and HKU-UCAS Joint Institute 
	for Theoretical and Computational Physics at Hong Kong, 
	The University of Hong Kong, Hong Kong, China}
\affiliation{The University of Hong Kong Shenzhen Institute of Research and Innovation, Shenzhen 518057, China}
\author{Gang Chen}
\email{gangchen@hku.hk}
\affiliation{Department of Physics and HKU-UCAS Joint Institute 
	for Theoretical and Computational Physics at Hong Kong, 
	The University of Hong Kong, Hong Kong, China}
\affiliation{The University of Hong Kong Shenzhen Institute of Research and Innovation, Shenzhen 518057, China}

\date{\today}

\maketitle

\section{Details of the first-principles calculations}

To obtain the hopping parameters of \ce{Sr2IrO4}, we use first-principles calculations to get the Kohn-Sham (KS) states and the eigenenergies. Then the maximally localized Wannier functions (MLWFs)~\cite{Marzari1997,Souza2001} are constructed, and the tight-binding (TB) model is established. The first-principles calculations are based on density functional theory~\cite{Hohenberg1964,Kohn1965} and plane-wave pseudopotential method, as implemented in the {\sc Quantum ESPRESSO} package~\cite{QE-2009,QE-2017}. The low temperature experimental crystal structure~\cite{Huang1994} is adopted in the calculations. The energy cutoff of the plane wave basis set is 120~Ry. The Brillouin zone is sampled by a $ 6 \times 6 \times 6 $ $ \vec{k} $-grid. The Schlipf-Gygi norm-conserving pseudopotentials~\cite{Schlipf2015} produced by the ONCVPSP code~\cite{Hamann2013} are used. The MLWFs and the TB model are constructed by the RESPACK program~\cite{Nakamura2021}. The initial guesses of the Wannier functions (WFs) are the projections of the KS states near the Fermi level to the local $ t_{2g} $ orbitals of each \ce{Ir} atom. The spatial spread of the WFs are then minimized.

\begin{figure}
	\includegraphics[width=10cm]{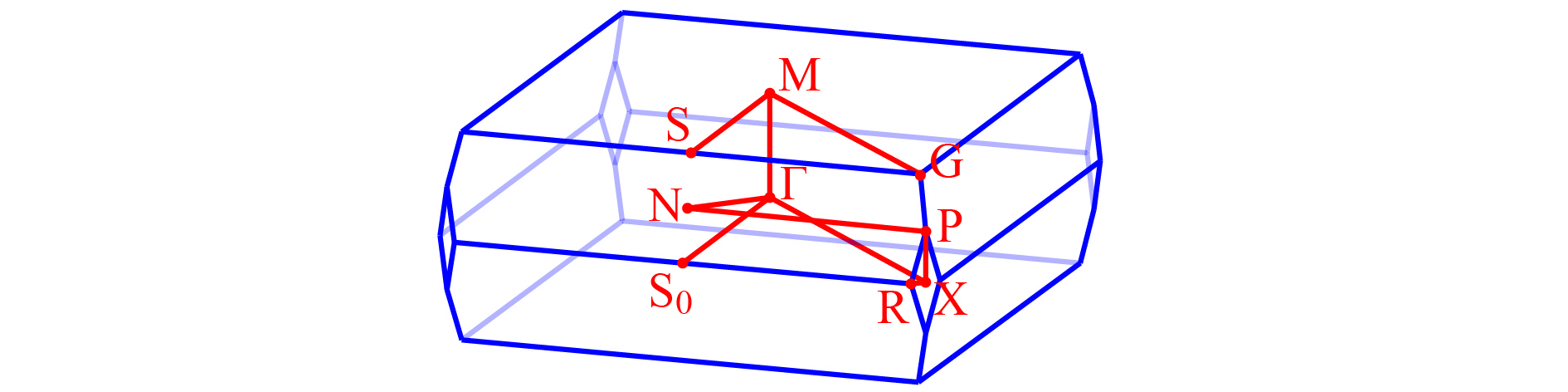}
	\caption{The Brillouin zone and the high symmetry paths of \ce{Sr2IrO4}.}
	\label{fig: BZ}
\end{figure}

\begin{figure}
	\includegraphics[width=12cm]{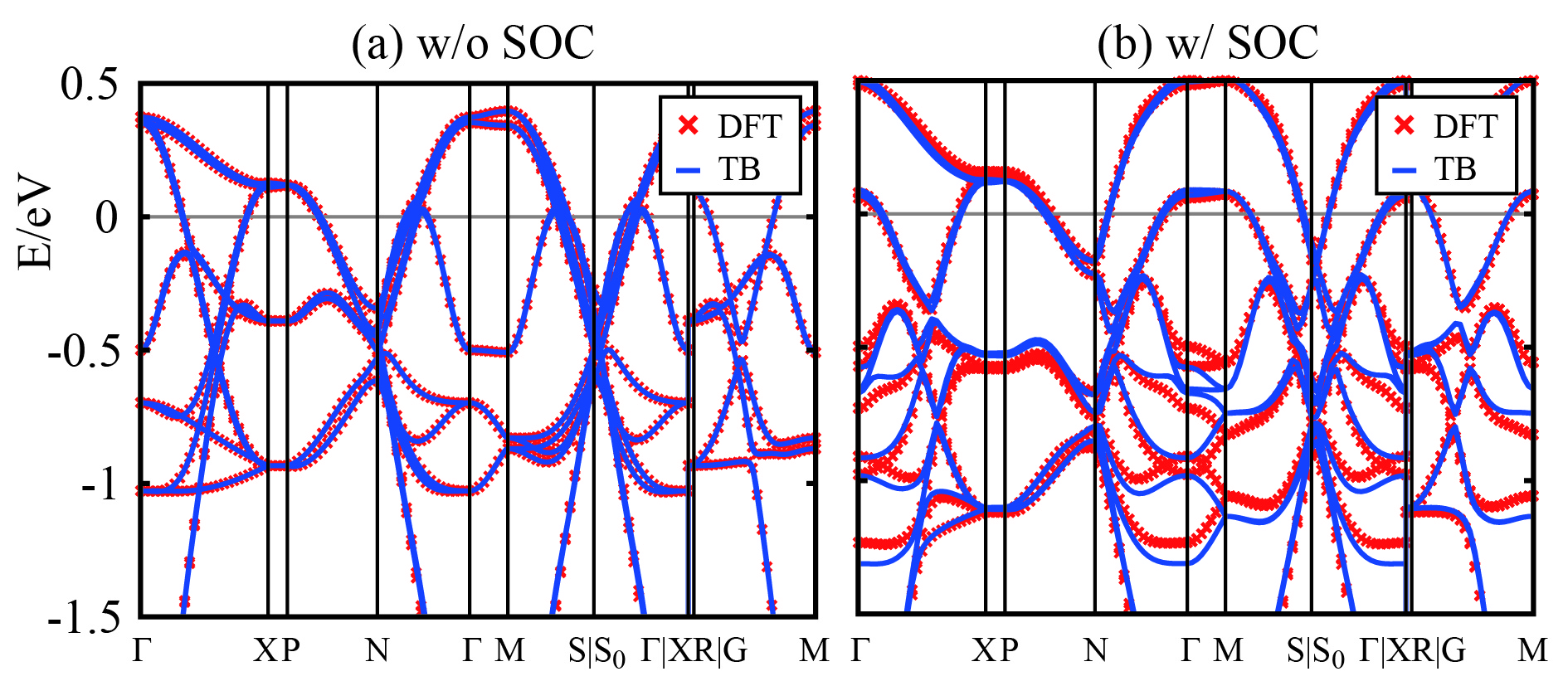}
	\caption{Comparison of the band structures calculated by DFT and the TB model, without SOC (a) and with SOC (b).}
	\label{fig: TB_fitting}
\end{figure}

The comparison of the band structures calculated by the TB model and DFT without spin-orbit coupling (SOC) is presented in \Fig~\ref{fig: TB_fitting}(a). The high symmetry paths in the Brillouin zone shown in \Fig~\ref{fig: BZ} are generated by the SeeK-path tool~\cite{Hinuma2017,Togo2018}. It can be seen that the tight binding model reproduces the DFT results perfectly. The nearest-neighbor and next-nearest-neighbor hopping parameters are listed in Table~I of the main text.

We then add to the obtained TB model a local SOC term (the second term of \Eq~(1) in the main text), and compare its band structure with the DFT result including SOC. The SOC strength $ \lambda $ is fitted by the least square method to be $ \lambda = 0.36~\text{eV} $. The corresponding band structures are shown in \Fig~\ref{fig: TB_fitting}(b). The details near the Fermi level ($ E=0 $) are well reproduced by the TB model with the local SOC term.

To evaluate the Hubbard parameter $ U $, constrained random phase approximation~\cite{Aryasetiawan2011} is adopted, as implemented in the RESPACK program~\cite{Nakamura2021}. The Hilbert space is divided into the subspace of the $ t_{2g} $ orbitals in which the MLWFs are constructed, and the rest subspace ($ r $ subspace). $ U $ is taken to be the orbital-averaged on-site static Coulomb interaction partially screened by the polarization processes involving the $ r $ subspace. It is found that $ U = 2.232~\text{eV} $.

\section{analytical expressions for the spin interaction parameters}

The analytical expressions for the spin interaction parameters of \ce{Sr2IrO4} in the main text are
\begin{eqnarray}
J&=&\frac{4\left[(t_{1}+t_{2}+t_{3})^{2}-4t_{4}^{2}\right]}{9U},\\
D&=&\frac{16(t_{1}+t_{2}+t_{3})t_{4}}{9U},\\
K&=&\frac{32t_{4}^{2}}{9U},\\
J^{\prime}&=&A\times\left[(t_{1}+t_{2}+t_{3})(-t_{1}+2t_{2}-t_{3})+4t_{4}^{2}\right],\\
D^{\prime}&=&A\times2t_{4}(2t_{1}-t_{2}+2t_{3}),\\
K^{\prime}&=&A\times8t_{4}^{2},
\end{eqnarray}
where the parameter
\begin{equation}
A=\frac{32(\lambda+U)(2t_{1}^{\prime}+t_{2}^{\prime}+t_{3}^{\prime})}{9U(3\lambda+2U)^{2}}.
\end{equation}
The corresponding numerical values are listed in Table~I of the main text.

\section{Effects of the tetragonal crystal field}

In deriving the above analytical expressions, we have assumed a perfect octahedral environment of the \ce{Ir^{4+}} ion. As a result, the three $ t_{2g} $ orbitals are degenerate. In the crystal of \ce{Sr2IrO4}, their is a tetragonal crystal field which makes the energy of the $ d_{xy} $ orbital different from that of the degenerate $ d_{xz} $ and $ d_{yz} $ orbitals. It gives rise to an additional term in the atomic Hamiltonian
\begin{equation}
H_{0i}^{\text{CEF}}=\sum_{\sigma}C_{i\sigma}^{\dagger}\left[\begin{array}{ccc}
	\Delta & 0 & 0\\
	0 & \Delta & 0\\
	0 & 0 & 0
\end{array}\right]C_{i\sigma},
\end{equation}
where $ C_{i\sigma}^{\dagger}=[c_{i,yz,\sigma}^{\dagger},c_{i,zx,\sigma}^{\dagger},c_{i,xy,\sigma}^{\dagger}] $ as in the main text. It is found that $ \Delta = 159~\text{meV} $.

There are no simple analytical expressions for the spin interaction parameters for nonvanishing $ \Delta $. The numerical results are listed in Table~\ref{tab: spin interation parameters for nonzero Delta}. Compared with the corresponding parameters in the Table~I of the main text, it can be seen that the tetragonal crystal field has only minor influences on the spin interactions.

\begin{table}
	\begin{ruledtabular}
		\begin{tabular}{cccccc}
			$ J $ & $ D $ & $ K $ & $ J^{\prime} $ & $ D^{\prime} $ & $ K^{\prime}$\\
			$ 59.3 $ & $ 14.2 $ & $ 1.67 $ & $ 2.56 $ & $ 0.597 $ & $ 0.209 $
		\end{tabular}
	\end{ruledtabular}
	\caption{The parameters of the spin interaction between site 1 and site 2 in \Fig~3(b) of the main text, in meV. The nonvanishing tetragonal crystal field is taken into account.}
	\label{tab: spin interation parameters for nonzero Delta}
\end{table}

\bibliography{refs_SM.bib}